%% file: main.tex
\documentclass[
reprint, 
superscriptaddress, 
amsmath,
amssymb, 
aps,
floatfix,
longbibliography
]{revtex4-1}

\usepackage{graphicx}
\usepackage{dsfont}
\usepackage{xcolor}
\usepackage[colorlinks=true,citecolor=blue,linkcolor=magenta]{hyperref}
\usepackage{bm}

\begin{document}

\title{Topologically protected Grover's oracle for the partition problem}

\author{Nikolai A. Sinitsyn}
\affiliation{Theoretical Division, Los Alamos National Laboratory, Los Alamos, New Mexico 87545, USA}
\author{Bin Yan}
\affiliation{Theoretical Division, Los Alamos National Laboratory, Los Alamos, New Mexico 87545, USA}

\begin{abstract}
   The Number Partitioning Problem (NPP) is one of the NP-complete computational problems. Its definite exact solution generally requires a check of all $N$ solution candidates, which is exponentially large. Here we describe a path to the fast solution of this problem  in $\sqrt{N}$ quasi-adiabatic quantum annealing steps. We argue that the errors due to the finite duration of the quantum annealing can be suppressed if the annealing time scales with $N$ only logarithmically. Moreover, 
   our adiabatic oracle is topologically protected, in the sense that it is robust against small uncertainty and slow time-dependence of the physical parameters or the choice of the annealing protocol. We also argue that our approach can solve many other famous NP-complete computational problems in $\sqrt{N}$ steps. 
   % This corresponds  to an exponential speedup in computation time.
   % We argue that our quantum computation strategy is considerably simpler to  achieve in practice than the universal gate-based quantum computation. 
\end{abstract}

\maketitle

\section{Introduction}

% There have been considerable advancements in the hardware for quantum information processing, especially in quantum state lifetime and fidelity of quantum gates. However, there are additional problems that are to be solved to enable practically useful quantum computing. 

%On the theoretical side,
%Thus,
The basic quantum algorithms, such as by Grover \cite{Grover1997}, matrix inversion \cite{m-inversion}, and the solution of the glued-trees problem \cite{tree},
assume that a  part of a targeted problem is pre-solved.  That is, such algorithms assume that a certain quantum function that points to the solution indirectly or a Hamiltonian that encodes the original mathematical problem is  given almost for free, i.e., can be {\it called as an  oracle}. In practice, the oracle is a quantum operator that is usually hard to construct. 

For a realistically interesting computational problem to benefit from such quantum algorithms, there must be a separate fast algorithmic and hardware implementation of its oracle, which is usually an unsolved problem. On the other hand, 
there are no
examples of provable scalable quantum speedups using quantum annealing in the oracle-free context. There are actually theoretical works arguing no quantum  speedup by a quantum annealing search for the ground state of an Ising spin Hamiltonian without considerable  symmetries in the problem  \cite{qa-bad,Yan2022Analytical}. 

Recently, the physical Grover's oracle implementation was suggested \cite{PRXQ-PP} for 
a solution of the Number Partitioning Problem (NPP)  \cite{PP-hard}. The idea in Ref.~\cite{PRXQ-PP} was to use resonant interactions of the computational qubits with a central quantum system (a spin or a photon). The state of the qubits that was to be marked by the oracle was interacting with a central system at resonance, so that the phase of this special state changed by $\pi$, while minimizing unwanted effects on the amplitudes of the other computational basis states. 

However, the resonant interactions with a targeted state are highly sensitive to the precision of the resonance conditions. Any uncontrollable mismatch of interactions or a small imperfection of the control pulses produces a proportional effect on the quantum state. On the other hand, the solution of an exponentially hard problem by the Grover algorithm requires an exponentially large number of the oracle calls, so that by the end of the algorithm any uncontrolled error is magnified by a factor $\sqrt{N}$, where $N=2^n$ and $n$ is the number of computational qubits. To eliminate such errors, we must set the coupling parameters and control fields in the system with the corresponding exponentially high  precision.

Thus, the Grover's speedup in Ref.~\cite{PRXQ-PP} for the computation time was achieved at the expense of another physical resource, which  was the precision of the physical coupling parameters and the control fields. We also note that for NPP such a trade of resources is known even for classical computing.
Thus, there are classical dynamic programming algorithms that achieve the exact solution of NPP in time $T\sim2^{n/2}$, just as with the Grover algorithm  but using exponentially large memory space, i.e., $\sim 2^{n/4}$ classical bits of memory \cite{PP-espace}. 
In the case of a quantum computer this exponential memory resource is not used, i.e., we deal with $O(n)$ computation space but the requirement of the exponentially high precision on the physical parameters is undesirable as well. 

A more specific problem with
the approach in Ref.~\cite{PRXQ-PP} is that its oracle affects the phases of the nonresonant states. Only in the adiabatic limit, these unwanted phases become truly suppressed, according to Ref.~\cite{PRXQ-PP}, as $\sim 2\arctan(E\tau_O)+\pi$, where $\tau_O$ is the duration of the interaction that generates the oracle and $E$ is the characteristic energy gap to the states that represent wrong solutions. Indeed, for $|E|\tau_O \gg 1$ such phases become close to either $0$ or $2\pi$, which would mean no unwanted error. However, for finite $E$ and $\tau_O$, the deviation is of the order $1/(|E|\tau_O)$. Hence, in order to make this phase error scale as $\sim 1/\sqrt{N}$, the time to produce the oracle has to scale as $\tau_O\sim \sqrt{N}$ at fixed $E$. Taking this into account, the entire time of the algorithm in Ref.~\cite{PRXQ-PP}  scales as $\tau_O\sqrt{N}\sim N$, which is the same as for the classical algorithm. Similar hidden costs can be found in other quantum algorithms, as we 
show briefly in Appendix~\ref{sec-QFT}.
This raises a question about whether such hidden costs on time and the trade of resources in quantum computing are inevitable.
 
In this article, we propose an approach  that essentially eliminates these hidden problems from the solution of NPP by the Grover algorithm during physical time $\sim \sqrt{N}$. 
Our approach  uses quasi-adiabatic quantum annealing in order to produce useful unitary transformations \cite{hen,adiabO}. Importantly, unlike Ref.~\cite{PRXQ-PP}, we do not  request knowledge of the precise position of the resonance with the searched state.  This makes our approach not only robust against the physical parameter uncertainty but also capable of solving a more complex version of NPP, as well as many other 
NP-complete problems that we will consider in Sec.~\ref{sec-gen}. %Instead, we use topologically protected $\pi$-phase shifts imposed on a range of the basis states that contain an unknown solution. We can then search for this solution by gradually decreasing this range. 

%The central spin type of interactions also encounters in quantum dots, with an electronic spin coupled usually to millions of nuclear spins \cite{q-dot} or a controllable number of magnetic ions \cite{crooker-cs}. Such interactions on the scale of $\sim10$ qubits also should be relatively easy to implement in ultra-cold atomic systems, which we hope will be used for a proof-of-principle experiment that will demonstrate our approach. 
%In what follows, we will not consider the hardware details any further -- we already request much less than the fully functional quantum computer to achieve our goals. 

\section{Number Partitioning Problem}

The NPP has the goal to split a
set ${\cal S}=\{s_1,s_2,\ldots,s_n\}$ of positive integers $s_k$, $k=1,\ldots,n$  into two subsets ${\cal S}_1$ and ${\cal S}_2$ so that the difference between the sum of integers in ${\cal S}_1$ and the sum of integers in ${\cal S}_2$ is minimized. There are different formulations of this problem. We will restrict ourselves here to its two specific versions that we will call NPP1 and NPP2. 

(i) In NPP1, the difference may be always nonzero, so the goal is to find the partition that delivers the minimal, in absolute value, difference between the two sums. 

(ii) In NPP2, it is assumed that the difference between the sums in ${\cal S}_1$ and ${\cal S}_2$ is known to be zero for some partitions, so the goal is to find at least one of them.

Both problems can be formalized by introducing $n$ binary variables $\sigma_k^z=\pm 1$ that mark the number $s_k$ as belonging to ${\cal S}_1$ if $\sigma_k^z=1$ and as belonging to ${\cal S}_2$ if $\sigma_k^z=-1$. 
NPP1 then has the goal to find components of an $n$-vector  $(\sigma_1^z,\ldots,\sigma_n^z)$ that provide the minimum,
\begin{equation}
 {\rm min}\left|H_I\right|,
    \label{cond2}
\end{equation}
where $H_I$ is a linear form
\begin{equation}
 H_I\equiv   \sum_{k=1}^n s_k\sigma_k^z.
 \label{hi-def}
\end{equation}

NPP2 is equivalent to finding the binary variables that satisfy a constraint
 \begin{equation}
H_I\equiv\sum_{k=1}^n s_k\sigma_k^z =0.
    \label{cond1}
\end{equation}
We can interpret the linear form $H_I$ as a simple Ising Hamiltonian of $n$ quantum spins-1/2. So, the goal of NPP1 is to find the eigenstate with the minimal nonnegative eigenvalue of $H_I$, and the goal of 
NPP2 is to find an eigenstate of $H_I$ that corresponds to zero eigenvalue. 

Note that for NPP1, the $H_I$-energy of the searched state is not {\it a priori} known. This is why the strategy in Ref.~\cite{PRXQ-PP} cannot be applied to NPP1 directly. Also NPP2 is a special case of NPP1. However, we will treat NPP2 separately because the knowledge of the energy of the searched state can be used for a simpler strategy. 
% can be considerably simpler. 
%This task is nontrivial because this eigenvalue is neither maximal nor minimal for $H_I$. Each positive eigenvalue of $H_I$ has a negative counterpart, which reflects that the indices $1,2$ of the sets ${\cal S}_1$ and ${\cal S}_2$ can be prescribed to theses sets arbitrarily. This means that the zero eigenvalue of $H_I$ is at least doubly degenerate. For simplicity we will assume for SPP1 that there is only a single such a zero energy doublet for $H_I$.
%so it is more convenient for early experimental verification.
The following facts have been established about NPP previously. 

First, NPP is NP-hard \cite{PP-hard}. Therefore, it is generally exponentially hard to solve exactly. Although Monte-Carlo algorithms in many situations produce the solution in time that scales with $n$ polynomially, in the worst cases the needed time is exponential: $T\sim2^n$. Thus, if we are to solve such a problem definitely and exactly, given only polynomial in $n$ memory resources, there is no better way than to test all $2^{n-1}$ independent possibilities for different $n$-vector solution candidates. In what follows, we will be concerned with the goal to find such an exact solution with probability exponentially close to $1$. 

NPP is NP-complete \cite{PP-hard}. All other NP problems can be solved faster if one finds a fast universal algorithm to solve any of the  NP-complete problems.  

NPP can be formulated as a Quadratic Unconstrained Binary Optimization (QUBO) problem, whose goal is to find the minimum of a quadratic form of binary variables \cite{PP-physics}. Thus, the quantum Ising spin Hamiltonian
\begin{equation}
 H_Q=H_I^2   
\end{equation}
has all nonnegative eigenvalues, so the state with the minimal eigenvalue  can be found by standard means of quantum annealing. However, the price for this strategy would be the requirement to build an all-to-all interacting qubit network, which is difficult in practice. Even then, we have to  deal with the lack of a known annealing protocol that would definitely outperform the classical search for the ground state of $H_Q$ with arbitrary free parameters. So, we will discard this strategy.
%as  not mathematically justified for our goal.

Finally, for any positive eigenvalue of $H_I$ there is the same eigenvalue but with a negative sign, with corresponding eigenstates different by the flip of all computational spins. The range of possible eigenvalues of $H_I$ is also known: Since all $s_k$ are  positive, the highest and lowest eigenvalues are provided by the fully polarized qubit states: $E_{max}=-E_{min} = \sum_{k=1}^n s_k$. Since all $s_k$ are integers, we definitely know that there is at least a unit gap between any two different eigenvalues of $H_I$. This also means that there are no  energy levels in a finite vicinity of the fractional energy values, e.g., near $E=1/2$. 

\section{Solution strategy}
\label{s-alg}
%Quantum annealing is known to provide a unitary evolution that may be employed by quantum algorithms \cite{hen}. Our strategy along this path is also based  on our recent observation in \cite{adiabO} that quantum annealing can be used to produce the oracles for the Grover's algorithm. Unlike Ref.~\cite{adiabO}, here we are not searching for the ground state  but satisfying a constraint,  (\ref{cond2}) or (\ref{cond1}). Nevertheless, a similar strategy applies. 

Consider any superposition of eigenstates of $H_I$,
\begin{equation}
  |\psi\rangle= \sum_{s=1}^{N} a_s|s\rangle, \quad N\equiv 2^n.
    \label{any-psi}
\end{equation}
We will show that by a single annealing step, whose time scales only as $\sim \log^{\alpha} N$, where $\alpha=O(1)$, we can generate an oracle that changes the sign of all state amplitudes with $H_I$-energy below an arbitrarily prescribed energy level $E$. The infidelity of this oracle is exponentially small in $n$. We use this oracle to change the sign of the states with eigenvalues of $H_I$ in the range $(-1/2,E)$ by applying the oracle at level $E$ and then applying it at level $-1/2$. This flips the  sign of the amplitudes of all basis states in (\ref{any-psi}) with only nonnegative eigenvalues below $E$. 

Being able to flip the signs for the states in the range $(-1/2,E)$, one can employ the algorithm of amplitude amplification~\cite{Brassard1997,Brassard2000} to find a basis state within this range with nearly unit probability, in $\sim \sqrt{N}$ steps. Within this range, the relative probabilities for the basis states to be found are determined by their relative weights $|a_s|^2$. In Appendix~\ref{a-aa}, we review the basics of the Grover algorithm and amplitude amplification. Let the found eigenstate correspond to an 
eigenvalue $E_k$. We then reset 
$$
E\rightarrow E_k+1/2.
$$

The NPP1 protocol starts  with an equal superposition of all the computational basis, i.e., $a_s=1/\sqrt{N}, \forall s$ in (\ref{any-psi}). With an initial trial value of the energy threshold $E$, we then repeatedly apply the procedure described above to update its value. The range $(-1/2,E)$ will then be shrunk so that $E$ becomes the lowest nonnegative eigenvalue of $H_I$, and therefore, the target state is found. Since the initial state of the amplitude amplification is the equal superposition, each eigenstate from the desired energy range can be found with equal probability. On average, each step of resetting $E$ reduces the number of eigenvalues in the interval $(-1/2,E)$ by a factor $2$. Hence, the  algorithm takes only $\sim \log_2N$ cycles to obtain the result with a close to $1$ probability. It takes then $\sim \log_2N$ repetitions of the entire process to make the probability of a wrong solution exponentially small. 
This completes the algorithm up to the procedure that generates the oracles, which will be the main ``know-how" result of our work. 

For NPP2, we will provide a process that for any superposition (\ref{any-psi})
produces, after a single quantum annealing step, almost the same superposition but with the flipped sign for amplitudes of all states $|s_{{\alpha}}\rangle$ that correspond to the zero eigenvalue, $H_I|s_{\alpha}\rangle=0$.
Having this, the desired eigenstate is found by a conventional Grover algorithm in $\sim \sqrt{N}$ repetitions of the quantum annealing process. 

%Thus, up to logarithmic corrections, the time to solve NPP is determined by the number of steps of the Grover's algorithm, which scales as $\sim\sqrt{N} = e^{n/2}$.

\section{Generating oracles}
\subsection{Basic hardware requirements}

As in Ref.~\cite{PRXQ-PP}, the most complex part of  hardware that we request is  the Ising central spin interaction Hamiltonian of the form
\begin{equation}
H_{int} = r\sum_{k=1}^n s_k \sigma_k^z I_z,
    \label{centralI}
\end{equation}
where $s_k$ are integers  and $\sigma_k^z$ are the Pauli $z$-matrices acting in phase space of the computational qubits; $I_z$ is the projection operator for an ancillary spin, and $r$ sets the energy scale.
In what follows, we will set the Planck constant $\hbar =1$, as well as $r=1$, which makes both energy and time dimensionless. Our energy and time variables can be reconstructed in physical units by multiplying them by, respectively, $r$ and $\hbar/r$.

Unlike Ref.~\cite{PRXQ-PP}, our approach specifies that the central spin has size $I=1$. We will also assume that we have access to high-fidelity quantum gates for rotating all the spins/qubits by a fixed angle (single qubit resolution is not needed).

The interactions of the type (\ref{centralI}) with the central spin $I=1$ are encountered in real physical systems. For example, the electronic spin of an NV${^-}$ center in diamond has electronic spin-1, which is coupled to many nuclear spins-1/2 of $^{13}$C isotopes via dipole interactions \cite{elesko}. The direct interactions between the nuclear spins are negligible due to their small g-factors. When needed, the nuclear spins can be rotated by rf-pulses, while the electronic spin can be controlled by external magnetic fields or optically.

The physical effect on which our oracle generation relies essentially is the Robbins-Berry topological phase \cite{Robbins1994}, which we briefly review in Appendix~\ref{aBR}. 
This phase is generated when a unit spin, $\hat{{\bf I}}$, interacts with an adiabatically changing magnetic field, ${\bf b}(t)$, with the Hamiltonian
\begin{equation}
  H(t)={\bf b }(t) \cdot \hat{{\bf I}},
    \label{fieldB}  
\end{equation}
so that the spin starts at its zero projection on the initial field direction; the field remains finite during the evolution and ends up pointing in the opposite to its initial direction. In Fig.~\ref{fig:sphere}(a), the black arrow curve shows an example of a trajectory that the field direction leaves on a unit sphere.
At the end of the protocol, the spin is in the initial physical state with zero spin projection on the initial axis but its quantum state acquires a phase $\pi$ that does not depend on the time-dependent  ${\bm b(t)}$. This makes this phase topologically protected, including against weak nonadiabatic transitions.

\subsection{Grover's oracle for NPP1}

\begin{figure}[t!]
    \centering
    \includegraphics[width=\columnwidth]{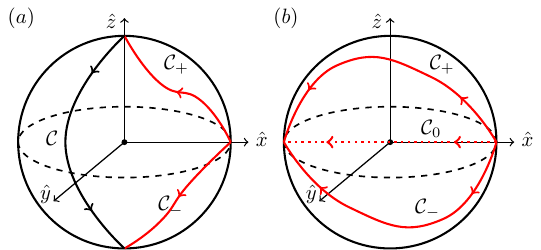}
    \caption{Paths of the adiabatically changing magnetic field direction $\bf{b}(t)/|\bf{b}(t)|$. The spin-$1$ is initially in the zero projection eigenstate along the field. It remains in the instantaneous zero-projection eigenstate during the time of evolution up to an accumulated phase. (a) The geometric phase along path $\mathcal{C}$, where the magnetic field flips its direction, is $\pi$. A closed path would generate no Berry phase ~\cite{Robbins1994}. Therefore, the phase difference between $\mathcal{C}_+$ and $\mathcal{C}_-$ is $\pi$. (b) The phases of paths $\mathcal{C}_\pm$ and $\mathcal{C}_0$ are, respectively, $\pi$ and zero.}
    \label{fig:sphere}
\end{figure}
For NPP1, we do not know {\it a priori} the energy of the state that we are searching for. Hence, we start with an arbitrary 
``guessed" value by generating a random eigenstate of $H_I$ and
measuring its eigenvalue $E_k$. If it is negative, we find the corresponding positive energy eigenstate by flipping all qubits. We  set the initial threshold to be $E=E_k+1/2$. 

\begin{figure}[t]
    \centering
    \includegraphics[width=\columnwidth]{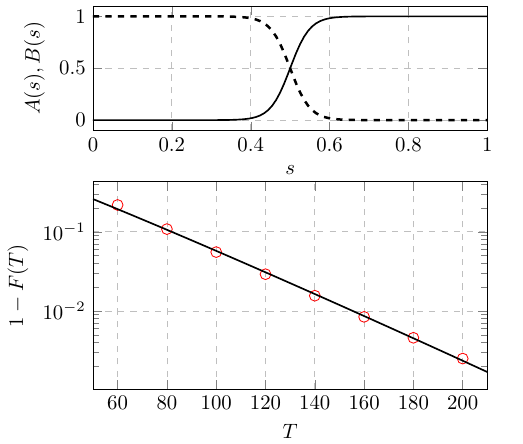}  \caption{Top: Annealing schedule in (\ref{gexp}), where the constant $c$ is fixed at $10$. Bottom: Simulation of the infidelity of the adiabatic oracle as a function of the total annealing time, for NNP1 with the problem set $\mathcal{S} = \{0,1,2\}$. The energy threshold for the oracle is set at $E=1.5$. Red circles are the numerical data. Black curves are the best fit to an exponential function $\sim\exp{(ax^b)}$ with $b\approx 1.08$.}
    \label{fig:NPP1}
\end{figure}

Then, we mark the amplitudes of all states that have energy $E_m<E$ by performing the quantum annealing  with the Hamiltonian 
\begin{equation}\label{Ha1}
 H_{a1}(s)= A(s) \left[\left(\sum_{k=1}^n s_k\sigma_k^zI_z\right)-E I_z \right] + B(s)I_x,
\end{equation}
where $I_{x(z)}$ is the spin-1 projection operator on the $x(z)$ axis. $s=t/T \in [0,1]$ is a dimensionless parameter, $t$ is time, and $T$ is the total annealing time. The annealing schedule, $A(s)$ and $B(s)$, is designed such that 
\begin{equation}
        A(0)=B(1)=0,\quad A(1)= B(0)= 1.
        \label{bound}
\end{equation}
%and at some intermediate time intervals $A(s)/B(s)\gg 1$.
The precise shape of the annealing schedule is not important, and we use the word ``adiabatic" in the sense that the evolution takes finite time but it is slow enough to suppress nonadiabatic excitations beyond some desired tolerance level. An example for the shapes of $A(s)$ and $B(s)$ are plotted in Fig~\ref{fig:NPP1}(top). We will discuss more precisely the requirements for the annealing schedules in Sec.~\ref{sec:fidelity}.

%Since the entire field is always nonzero, designed such that nonadiabatic excitations are suppressed exponentially fast in the total annealing time $T$, as will be discussed in the next section.

The annealing Hamiltonian $H_{a1}$ in  (\ref{Ha1}) is trivially solvable because it does not contain the terms that  flip computational qubits. Since $H_{a1}$ commutes with time-independent $H_I$, the evolution with $H_{a1}$ splits into $N$ invariant 3$\times$3 sectors, with the $k$th sector  corresponding to a conserved eigenvalue $E_k$ of $H_I$. Within this sector, the effective Hamiltonian  $H_{a1}$ has the form 
\begin{equation}
 H_{k}(s)=A(s)(E_k-E) I_z+B(s)I_x.
    \label{heffPP1}
\end{equation}

The evolution starts 
with the state that is a direct product of an arbitrary superposition $|\psi\rangle$ of states of the computational qubits and the zero projection state of spin ${\bm I}$ on the $x$-axis:
\begin{equation}
    |\Psi\rangle = |\psi \rangle \otimes |0_x\rangle.
    \label{init-s}
\end{equation}
The spin-1 state $|0_x\rangle$ is the eigenstate of the initial $H_{1a}$ at $s=0$. During the adiabatic evolution, in each sector the spin follows the instantaneous zero-projection state
$|0_{{\bf b}_k(s)}\rangle$
on the direction of the effective field with components ${\bf b}_k(s)\equiv (b_x,b_y,b_z)=\left(B(s),0,A(s)(E_k-E)\right)$.  The corresponding eigenvalue of $H_k$ in each sector is identically zero: $H_k(t)|0_{{\bf b}_k(t)}\rangle=0$. Hence, the dynamic phase is not generated.

In Fig.~\ref{fig:sphere}(a) we show that for $E_k>E$, the direction of ${\bf b}(t)$ changes from the direction of the $x$ axis to the direction of the $z$ axis. For $E_k<E$, however, the field ends up pointing in the opposite to the $z$ axis direction. In either case, the central spin ends up in the zero projection state, $|0_z\rangle$, on the $z$ axis.
However, the difference between the geometric phases generated by these two paths [red arrow curves in Fig.~\ref{fig:sphere}(a)] is the same as the phase generated by the field that switches  from the positive  to the negative direction along the $z$ axis.  
According to Ref.~\cite{Robbins1994} (see also Appendix~\ref{aBR}), this leads to an acquired  topological $\pi$-phase difference between the sectors with $E_k-E> 0$ and $E_k-E<0$. 

Summarizing, if  the initial state before the annealing is 
\begin{equation}
|\Psi_{in} \rangle =\left(\sum_k a_k |k\rangle \right) \otimes|0_x\rangle,
\label{init-s2}
\end{equation}
then after the annealing the state is 
\begin{equation}
|\Psi_{out} \rangle =\left(\sum_k (-1)^{\delta(k)}a_k |k\rangle \right) \otimes |0_x\rangle,
\label{oracle1}
\end{equation}
where $\delta(k)=1$ for $E_k<E$ and $\delta(k)=0$ for $E_k>E$, as it is required for the solution of NPP1 described in Sec.~\ref{s-alg}.

\subsection{Grover's diffusion step}

In addition to Grover's oracle, the Grover algorithm employs a Grover's diffusion step, which is an application of a unitary operator

\begin{equation}
U_{\rm GD}=2|\Rightarrow \rangle\langle \Rightarrow|-\mathds{1},
    \label{gd}
\end{equation}
where $|\Rightarrow \rangle$ is the state with all computational spins-1/2 rotated to point along the $x$ axis, and $\mathds{1}$ is the unit operator. While formally this step can be performed with a polynomial number of gates, as in Ref.~\cite{PRXQ-PP} we can generate it with a similar annealing step. 

Note that $U_{\rm GD}$ has the same structure as the Grover's oracle in the sense that $U_{\rm GD}$ merely changes the relative sign of the amplitude of a particular state of the computational qubits. The only problem is that this state, $|\Rightarrow \rangle$, is not an eigenstate of $H_I$. However, if we have an access to a unitary operator 
\begin{equation}
U_{\rm GDz}\equiv 2|\Uparrow \rangle\langle \Uparrow|-\mathds{1},
    \label{gdz}
\end{equation}
where $\Uparrow$ is the fully spin-polarized state along the $z$ axis,
then a simple  rotation of all spins from the $z$ axis to the $x$ axis direction transforms $U_{\rm GDz}$ into $U_{\rm GD}$. If all computational spin qubits are identical, this unitary operation is achieved with a simple pulse of a magnetic field:
\begin{equation}
U_{rot}=e^{-i(\pi/4) \sum_{k=1}^n \sigma_k^y},
    \label{urot}
\end{equation}
so that 
$$
U_{\rm GD}=U_{rot}U_{\rm GDz}U_{rot}^{\dagger}.
$$

The Hamiltonian $H_I$ has a nondegenerate state with all spins polarized along $z$-axis, which corresponds to $H_I$ eigenvalue $E_{max}=\sum_{k=1}^{N}s_k$. Since the energy of this state is known, we can mark amplitudes of all other states with a $-1$ factor by setting $E=E_{max}-1/2$ and performing a single annealing step. 
Thus, we do not have to change the interaction part of the Hamiltonian: the diffusion step is achieved with the annealing step as for the Grover oracle but in a different  field acting on the ancillary spin. %Moreover, we do not have to change the state of the ancillar qubit or adjust the $x$-component of the field after producing the Grover's oracle, as the time-reversed protocol, $g(-t)$, produces the same effect on the final state.

The application of the spin rotation  before and after this annealing with (\ref{Ha1})  produces the  equivalent effect to the application of the Grover's diffusion operator. The fact that no other quantum gates are needed is practically useful because a simple spin rotation can be performed with very high fidelity, e.g.,  $\sim 10^{-6}$ \cite{gate-high} probability of the error, whereas the entire universal set of quantum gates  cannot be usually produced with the fidelity better than $\approx 99\%$. What is  important for our discussion is that such a rotation of spin qubits can be done by rotating the control field quasi-adiabatically. The  precision  and time-scaling of this process then is not worse than for the oracle generation.

 \subsection{Fidelity of the oracle} \label{sec:fidelity}
In Grover algorithm, the oracle is called $\sim \sqrt{N}$ times, so it is required that the error does not accumulate to $O(1)$ probability of a wrong state after $\sqrt{N}$ annealing steps. This imposes a constraint on the tolerance of the nonadiabatic excitations and the running time of the adiabatic oracle.

With suitable time-dependent annealing schedules, one can suppress non-adiabatic deviations exponentially in the total running time $T$ \cite{Lidar2009Adiabatic,Ge2016Rapid,Albash2018Adiabatic}. Generally, the nonadiabatic errors scale as
\begin{equation}
    P_{\rm ex} \sim \exp{(-\eta \Delta^2/\beta)},
\end{equation}
where $\eta$ is a numerical factor depending on the specific annealing schedule, $\Delta$ is the characteristic gap near an avoided crossing point and $\beta$ is the rate of the transition through this gap. In our case, the lowest gap is found in the sector with $\Delta = |E-E_k|=1/2$.

An example of the protocol with exponential suppression of the errors is 
\begin{equation}\label{gexp}
 A(s), B(s) = \frac{1}{2}\left[ 1 \pm \tanh{c(2s-1)} \right],
\end{equation}
where $c$ is a large constant to ensure that the annealing schedule starts and terminates smoothly (derivatives of the schedules are suppressed~\cite{Lidar2009Adiabatic}). Note that if $c$ is of the order of $n=\log_2N$, the deviations of the boundary values of $A(s)$ and $B(s)$ from (\ref{bound}) are exponentially small. Therefore, we ignore errors caused by the imperfect boundary condition of the annealing schedules. Shapes of $A(s)$ and $B(s)$ are plotted in Fig.~\ref{fig:NPP1}(top). 

To quantify the accuracy of the oracle with the above annealing schedule, we simulated its infidelity as a function of $T$. The infidelity is defined as the $1-F(T)$, where $F(T)$ is the probability for the final output state of the oracle to be detected in the desired output state of an ideal oracle. In Fig.~\ref{fig:NPP1}(bottom), the exponential decay of the infidelity is observed.

Since the oracle is called $\sim \sqrt{N}$ times, the error of each oracle call must scale as
\begin{equation}
    P_{\rm ex} \sim 1/\sqrt{N},
    \label{supr}
\end{equation}

For our protocol, the rate of the transition through the gap is $\beta \sim c/T$. Since, $c \sim n$, the condition (\ref{supr}) is satisfied if 
$e^{-\eta T/n} \sim 2^{-n/2}$ for some $\eta = O(1)$.
This condition implies that the running time of the oracle satisfies
\begin{equation}
    T \sim \log^{2}{N}, 
\end{equation}
which retains the overall quadratic speedup of the Grover algorithm.

%%%%%%%%%%%%%%%%%%%%%%%%%%%%%%%%%%%%%%%

\section{Simpler approaches}

In this section, we discuss possible strategies to simplify experimental verification of our approach.
First, one can reduce the number of steps by considering the NPP2 version of the problem, in which the target state of the corresponding Ising Hamiltonian $H_I$ is known to have zero energy. This knowledge can be used to simplify the generation of the oracle. We will then discuss a strategy that does not involve time dependent tuning of the interaction strength between the Ising spins. This may be important for experiments without access to time-dependent interactions.

\subsection{Simplified oracle for 
NPP2}\label{ssec-NPP2}

\begin{figure}[t]
    \centering
    \includegraphics[width=\columnwidth]{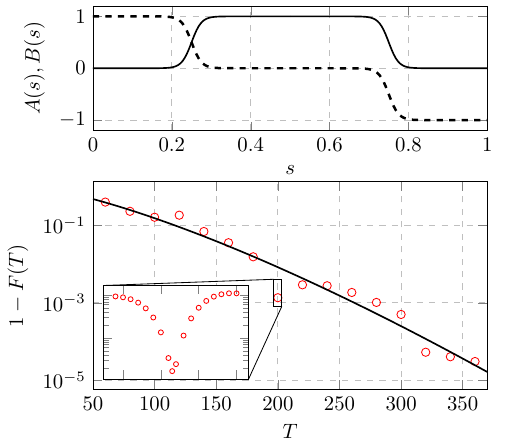} \caption{Top: Annealing schedule in (\ref{gt1}), where the constant $c$ is fixed at $10$. Bottom: Simulation of the infidelity of the adiabatic oracle as a function of the total annealing time, for NNP2 with the problem set $\mathcal{S} = \{1,2,3\}$. Red circles are the numerical data. Black curves are the best fit to an exponential function $\sim\exp{(ax^b)}$ with $b\approx 1.36$. Inset shows a zoom in the oscillation of the nonadiabatic excitation probability on top of the overall exponential decay.}
    \label{fig:NPP2}
\end{figure}

For NPP2, the $H_I$-energy of the searched state is known: $E_0=0$. Since this state belongs to the energy range $(-1/2,1/2)$, we can generate its Grover's oracle by performing annealing with the Hamiltonian $H_{a1}$ initially at $E=1/2$ and then at $E=-1/2$. Note that, as for NPP1, this approach is topologically protected. Namely, the physical parameters $s_k$ can be set not precisely and even can experience  slow time-dependent deviations from the desired integer values. Nevertheless, the topological $\pi$-phase is robust as long as the level $E$ is set in the gap that separates the searched state from the other states.

If the zero energy of the  searched state is protected by symmetry of interactions, the oracle for NPP2 can be generated in only a single quantum annealing step with the time-dependent Hamiltonian
\begin{equation}
  H_{a2}(s)= A(s)\left(\sum_{k=1}^{N}s_k\sigma_k^zI_z\right) +B(s)I_x,
   \label{ha2}
\end{equation}
where $A(0)=A(1)=0$, and $B(0)=-B(1)=1$. For example, such an annealing protocol can be created by combining the schedules in NPP2: 
\begin{equation}\label{gt1}
     A(s), B(s) = \begin{cases} 
    \frac{1}{2} \left[ 1 \pm \tanh{c(4s-1)} \right]  & s \le 1/2,\\
   \pm \frac{1}{2} \left[ 1 - \tanh{c(4s-3)} \right]  & s > 1/2.\\
    \end{cases}
\end{equation}
The shape of this schedule is plotted in Fig.~\ref{fig:NPP2}(top), in which we also demonstrate that nonadiabatic errors of this oracle are suppressed with the total annealing time $T$ exponentially.

The corresponding effective magnetic field ${\bf b}(s)$ switches direction to the opposite one by the end of the annealing, as we illustrate in Fig.~\ref{fig:sphere}(b). According to Ref.~\cite{Robbins1994}, this leads to the same state $|0_x\rangle$ at the end of annealing as at the beginning but with an acquired  topological $\pi$-phase in all sectors with $E_k\ne 0$. 

In contrast, for the eigenstates of $H_I$ with the eigenvalue $E_0=0$, the state $|0_x\rangle$ remains the exact eigenstate of the time-dependent Hamiltonian $H_0(s)=B(s)I_x$ with zero eigenvalue. Hence, during the entire protocol this state does not change and does not even acquire any dynamic or geometric phases.

Summarizing, if  the initial state before the annealing is (\ref{init-s2})
then after the annealing the state is 
\begin{equation}
|\Psi_{out} \rangle =\left(\sum_k (-1)^{\delta(k)}a_k |k\rangle \right) \otimes |0_x\rangle,
\label{oracle1}
\end{equation}
where $\delta(k)=1$ for $E_k\ne 0$ and $\delta(k)=0$ for $E_k=0$.
%This completes our construction of the oracle that creates an additional $\pi$-shift between the searched states in NPP1 with zero energy and the rest of the spectrum. 

\subsection{Annealing with time-independent couplings}
A caveat of the standard annealing schedule discussed above is that the time dependent $A(s)$ appears in front of the coupling terms of the Ising spins. Experimentally, changing the interaction strength could be hard to achieve, e.g., if the computational qubits are nuclear spins. Here, we introduce an annealing protocol with fixed coupling strengths.
For NNP1, in contrast to (\ref{Ha1}), the oracle is realized with
the Hamiltonian 
\begin{equation}\label{hag1}
 H'_{a1}(t)= \left(\sum_{k=1}^n s_k\sigma_k^zI_z\right)-E I_z+g(t)I_x,
\end{equation}
where the time changes in the interval $t\in(T_{\rm min},T_{\rm max})$ such that 
$$
g(T_{\rm min}) \gg 1, \quad g(T_{\rm max}) \ll 1.
$$
An example of such a protocol is 
\begin{equation}
  g(t)=e^{-t/T},  
  \label{expg1}
\end{equation}
where $T_{\rm min} \sim -T n$ and $T_{\rm max} \sim T n$. 

Considering no environmental decoherence, the errors for this oracle originate from two sources: (i) the finite time of the evolution, which leads to the nonadiabatic transitions over the energy gap and (ii) the finite interval of the external field values $g(t)$, which leads to the error $\sim |E_k/g(T_{\rm max})|$ due to misalignment of the initial field ${\bf b}$ from the $x$-axis. 

Given that nonzero eigenvalues $E_k$ of $H_I$ are integer numbers, the adiabatic conditions correspond to $T\gg 1$ in order to guarantee that in the worst case with  $|E_k-E|=1/2$ we avoid the nonadiabatic transitions during the evolution within each $E_k$ sector. In Appendix~\ref{a-nonad} we calculate the nonadiabatic transition probability for the protocol (\ref{expg1}) analytically, and thus verify its exponential suppression with $T$.

%To suppress the nonadiabatic transitions, the entire time of the evolution should scale only quadratically with $n$: $T\sim \alpha n^2$, where $\alpha=O(1)$. 

For the Hamiltonian (\ref{hag1}),  the boundary-related errors are suppressed if
the physical interval for $g(t)$ is sufficiently large, so that at the beginning and the end of the evolution the deviation of the entire field from the $x$-axis direction is exponentially suppressed, e.g., 
\begin{equation}
    g(T_{\rm max}) =-g(T_{\rm min}) \sim e^{\eta n},
    \label{expg2}
\end{equation}
where $\eta>1/2$ is 
chosen to make sure that the boundary error is not accumulated substantially after
$\sqrt{N}$ calls of the oracle. This guarantees that we are able to prepare the initial state of the spin-1 in all sectors as the zero projection  on the $x$-axis eigenstate.   Note, however, that due to the exponentially fast 
changes of $g(t)$, the entire time of the field sweep depends on $n$ only linearly. So, the entire time of the annealing step still scales logarithmically with $N\equiv 2^n$:
$$
T_{\rm max}-T_{\rm min} \sim \log ^{\alpha} N, \quad \alpha=O(1).
$$
 
%Our goals can be equally well achieved with the annealing Hamiltonian 
%$$
%H_{a2}'=A(t)H_I\otimes I_z +B(t) I_x, 
%$$
%where $A(t)$  switches the interactions on and off smoothly,  e.g., as $A(t)=1/\cosh(t/T)$, and $B(t)\sim g(t)$ but  for a much smaller evolution time interval. Then, all our discussions apply equally well but the amplitude of  $B(t)$ is no longer required to be exponentially large. The values $B(T_{max})\sim-B(T_{min})\gg E_{up}$ are then sufficient to suppress the boundary error below the required level, as we illustrate numerically in Fig.~\ref{fig:NPP2}. 

The condition (\ref{expg1}) suggests that if the couplings are time-independent we still need a large resource in the form of an exponentially large interval for the range of $g(t)$. Experimentally, allowing no time-dependent control of the interactions may simplify the first demonstrations of our approach. However, we expect that the time dependent interactions will be required with growing $n$ in order to reduce the  range for the accessible external field. 

Finally, we note that the most complex instances of NPP are very rare unless the largest integer number in the set ${\cal S}$ is exponentially growing with $n$ \cite{PP-hard}. In such situations, our annealing protocols still keep the annealing time logarithmic, albeit with an extra power of $\log N$. However, the energy range for both spin-spin interactions and the external field has to grow with $n$ exponentially. This resource requirement, however, is inevitable if we are to encode exponentially large input values in physical parameters. A strategy to alleviate this problem can be found in Ref.~\cite{PRXQ-PP}.

\section{Generalization to many constraints}
\label{sec-gen}

Let us finally comment on possible extensions of our approach to more difficult constraints satisfaction problems.
If the energy range for the couplings in $H_I$ is restricted, the number of states that satisfy a single constraint is typically exponentially large. However, $m=O(n)$ independent constraints of the form 
\begin{equation}
    H_I^{(k)}=a_k,
    \label{cons1}
\end{equation}
or 
\begin{equation}
H_I^{(k)}\ge b_k, \quad k=1,\ldots,m,
\label{cons2}
\end{equation}
can be usually satisfied simultaneously by only $O(1)$  states, as e.g., in the graph coloring problem \cite{QUBO-rev}. 
This makes the multiple constraint satisfaction generally classically hard even when the coupling parameters are similar in size. Here $H_I^{(k)}$ are the linear forms of binary variables with integer coefficients. They are different for different $k$; $a_k$, $b_k$ are independent integers. 

Let $N_1,\ldots, N_m$ be the numbers of states that satisfy, respectively, the first, the  first two, and so on up to all $m$ such constraints, and let us introduce the ratios:
$$
n_s=\frac{N_{s-1}}{ N_s}, \quad s=1,\ldots,m, \quad {\rm where}\,\,\, N_0\equiv N.
$$
The $n_s$ should be possible to find using the quantum algorithm for the number of solutions estimate in $\sqrt{n_s}$ calls of the Grover's oracle for each $n_s$, without changing the leading scaling of the time of the entire algorithm that we now describe. %If the coefficients of the different constraints can be treated as statistically independent random integers, the numbers $n_s$ may be also estimated with high precision using classical Monte-Carlo sampling during a polynomial in $n$ time.
As our goal is only to demonstrate further research directions, here we assume that we deal with a problem, for which all $n_i$ are given to be known.

We can prepare the oracle for each constraint separately. So, let us start with the first constraint and use its oracle to implement the Grover algorithm. In $\sim \sqrt{n_1}$ oracle calls, we will thus prepare a state $|+\rangle_1$, which is the superposition of all $N_1$ states that satisfy the first constraint. 

Let us look at the preparation of the state $|+\rangle_1$ as at application of a unitary operator $U_1$, such that $|+\rangle_1=U_1|\Uparrow\rangle$. By reversing the sequence of our field pulses, we can create an operator $U_1^{-1}$ in $\sim \sqrt{n_1}$ steps with oracle calls. Thus we can use this operator for the amplitude amplification algorithm that creates an overlap between the initial state $|\Uparrow \rangle$ and the superposition state $|+\rangle_2$ of all states that satisfy the first two constraints. 

Note that $|\langle \Uparrow | U_1^{\dagger} |+ \rangle_2|^2 =1/n_2 $.
Hence, it takes $\sim \sqrt{n_2}$ calls of this unitary and its inverse, as well as the oracle that marks the states that satisfy the second constraint, in order to prepare $|+\rangle_2$ using the amplitude amplification. Thus, it takes totally $\sqrt{n_1n_2}$ steps with calls of the constraint-marking oracles in order to prepare this state from the initial $|\Uparrow \rangle$. 

We can then treat the preparation process of $|+\rangle_2$ as a unitary operator $U_2$ action, whose time to implement takes $\sim \sqrt{n_1n_2}$ more elementary steps. The construction of the state $|+\rangle_3$ would then take $\sim \sqrt{n_1n_2n_3}$ such steps and so on. By induction, we find that the preparation of the state that satisfies all $m$ constraints would take $\sim \prod_{k=1}^m\sqrt{n}_k=\sqrt{N/N_m}\sim \sqrt{N}$ calls of the fast oracles, whose construction we already described. Thus, the introduction of multiple constraints does not affect the $\sim \sqrt{N}$ scaling, at least for the ``typical" situations for which the numbers $n_i$ can be quickly estimated.

The number of constraints that can be satisfied is restricted by the error with which the Grover algorithm can prepare the sequence of states $|+\rangle_1 \rightarrow |+\rangle_2 \rightarrow \ldots \rightarrow |+\rangle_m$. For example, instead of $|+\rangle_1$, the algorithm prepares a state $|+\rangle_1+c_1|e_1\rangle$, where $e_1$ is some error state with amplitude $c_1 \sim 1/\sqrt{n_1}$. Iterating, we find that instead of the final $|+\rangle_m$, the algorithm prepares a state 
$$
|+\rangle_m+\sum_{k=1}^m c_k|e_k\rangle,
$$
where $c_k\sim 1/\sqrt{n_k}$. The error states $|e_k\rangle$ are generated from strongly different initial states, so they are expected to be essentially orthogonal to each other. Hence, altogether, they can be considered as a state orthogonal to $|+\rangle$ with an amplitude $\sim \sqrt{c_1^2+\ldots +c_m^2}$. For example, if all $n_k$ are of the order ${n}$, then $O(n)$ constraints produce an error with the probability comparable to the one of the correct result. This would still  be acceptable because the correct solution can be found then after $O(1)$ repetitions of the entire algorithm.  

Finally, in some of the QUBO problems, such as the set partitioning and minimum vertex cover problems \cite{QUBO-rev}, in addition to the constraints (\ref{cons1}) and (\ref{cons2}) we must minimize some linear form:
$$
{\rm find} \,\,\,{\rm min} \left(\sum_{k=1}^n d_k\sigma_k^z\right), 
$$
with integers $d_k$.

We already described how to prepare a unitary $U$, that transforms $|\Uparrow \rangle$ into the superposition $|+\rangle$ of all states that satisfy (\ref{cons1}) and (\ref{cons2}).
We can then use it with the oracle that marks all states in this superposition below arbitrary energy level $E$. The Grover algorithm then is used to produce the state that contributes to $|+\rangle$ and has a lower eigenvalue than $E$. This allows us to update $E$ as in Sec.~\ref{s-alg} (see also Refs.~\cite{Durr1996,adiabO} for similar approaches to energy minimization), and determine the solution of such a problem in a logarithmic number of the level $E$ updates.

%%%%%%%%%%%%%%%%%%%%%%%%%%%%%%%%%%%%%%%

\section{Discussion}
The NPP is one of the practically most useful famous computational problems. We showed that quantum mechanics allows its general exact solution with probability exponentially close to $1$ faster than the classical solution, and essentially without an exponential overhead due to the control precision. The computational memory is polynomial in the size of the partition problem. In contrast, many classical algorithms require exponential memory to achieve a speedup. 

The computation time $T_{\rm comp}\sim 2^{n/2}$ of our algorithm still scales exponentially with the number of integers that should be partitioned. However, this quadratic speed up may still provide a quantum advantage: %the quantum speedup is exponential in the sense that the computation time is reduced by an exponentially small factor in comparison to the classical $T_{\rm comp}\sim 2^{n}$. 
For modern classical computers, the exact solution of NPP should become generally impossible for $n\sim 60$, which corresponds to an order of $2^{60/2} \sim 10^{9}$ calls of the oracle in the Grover algorithm. Thus, we estimate that the quantum supremacy for this problem can be achieved if the quantum annealing model with the central spin interactions is implemented for $n\approx 60$ qubits, with $\sim 10^{-9}$ error rate per one annealing step. For the qubits with the quantum lifetime of order $1$s, these steps should take not more than $1$ns. Altogether this is still beyond the ability of modern quantum technology but the numbers are not too far away from what is possible. For example, similar estimates show that our approach is within the modern experimental reach for $n\approx 40$, which would be hard for a desktop. For such $n$, we need $\sim 10^6$ oracle calls, with the fidelity that was demonstrated in some systems \cite{gate-high}.

Finally, we comment on a recent work \cite{grover-bad}, claiming that the Grover algorithm provides no quantum advantage. The criticism in Ref.~\cite{grover-bad} was based on the assumption that the Grover's oracle is constructed as a separate quantum circuit. The authors in \cite{grover-bad} argued that for the cases when this circuit can be simulated classically, the problem is also solvable entirely by a classical computer. Hence, for many known classically complex problems, the Grover's oracle may be hard to implement as a quantum circuit. For example, it can be hard to design such an oracle using a classical computer. 

Our work does not contradict Ref.~\cite{grover-bad}. Namely, we do not know a short circuit that would simulate our quantum annealing step on a gate-based quantum computer with the desired accuracy. For example, the Suzuki-Tr\"otter decomposition requires  $\sim \sqrt{N}=2^{n/2}$ quantum gates in order to simulate our annealing step with accuracy $O(1/\sqrt{N})$, which would be needed to suppress the discretization errors throughout all $\sim \sqrt{N}$ steps of the Grover algorithm. Hence, our approach may not provide an advantage if it is implemented as a fully gate-based quantum circuit,
unless using methods designed to accelerate gate-based quantum annealing simulations
 \cite{signal}.

We showed, however, that this problem can be avoided with a physical quantum annealing evolution, which can be performed in time that scales  with $N$ only logarithmically and employs only simple interactions between qubits. Thus, we resolved the question in favor of quantum computers without arguing against the analytical results in Ref.~\cite{grover-bad}.

\begin{acknowledgements}
This work was supported in part by the U.S. Department of Energy, Office of Science, Office of Advanced Scientific Computing Research, through the Quantum Internet to Accelerate Scientific Discovery Program, and in part by U.S. Department of Energy under the LDRD program at Los Alamos.
\end{acknowledgements}

\input{appendix}

\bibliography{references}
\clearpage

\end{document}

%% file: appendix.tex
\appendix

\section{Hidden energy cost of Quantum Fourier Transform (QFT)}

\label{sec-QFT}

The QFT \cite{QFT} is a component of many quantum algorithms, such as the Shor's algorithm  \cite{shor}. It can be implemented with a polynomial in the number of qubits, $n$, basic quantum gates. However, its practical implementation in hardware contains a hidden exponentially growing cost, which is similar to the one that we discuss in Introduction. 

Namely, a basic requirement for the QFT is to use a controlled phase shift, associated with a unitary operator 
\begin{equation}
  R_k=\left(\begin{array}{cc}
  1 & 0 \\
  0 & e^{2\pi i/2^k}
  \end{array}
  \right).
    \label{Rn}
\end{equation}
The standard  estimate for the physical QFT algorithm performance assumes implicitly that such operators can be called in a finite time $\tau$ for all $k=1, \ldots, n$. In practice, however, such a phase shift is induced by 
switching on the coupling between the qubits during the time duration $\tau$ with the characteristic coupling energy 
$$
E_k = 2\pi /(\tau 2^k).
$$
Hence, the accessible energy bandwidth for this coupling has to range from $E_n \sim 1/(N\tau)$ to $E_1 \sim 1/\tau$, where $N=2^n$. 

Such an energy resource is hard to provide physically. For example, if  we assume that the qubit is rotated by an effective magnetic field that can be set in the range of 1 Tesla, which is $10^4$ Gauss, with the precision of only $1$ Gauss, then the number of matrices $R_k$ that we can implement in one time step is restricted by $n=\log_2 10^4\approx 13$, which is still too small for commercial applications. 

Moreover, the physical energy bandwidth for the qubit control is always finite, as well as our ability to discretize this bandwidth by distinct coupling energies.  Hence, as $n$ is growing, the gates $R_k$ have to  be composed generally of repeated applications of the gates from the finite subset of the readily accessible controlled phase shifts. Then, the time to implement  the QFT algorithm scales with $N=2^n$ linearly.

\section{Amplitude amplification}
\label{a-aa}

Given a quantum state in an equal superposition of $N$ basis states, i.e.,
\begin{equation}
    |s_0\rangle = \frac{1}{\sqrt{N}} \sum_{i=1}^N |i\rangle,
\end{equation}
Grover algorithm finds the target state $|\omega\rangle$ in $\sim \sqrt{N}$ steps. The basic ingredient of the Grover algorithm is the oracle operation:
\begin{equation}
    \hat{O} \equiv I-2|\omega\rangle\langle\omega|,
\end{equation}
which flips the sign of the target state and keeps the other basis states unchanged. Each oracle call is also supplemented by a diffusion operator, defined as
\begin{equation}
    \hat{D} \equiv I-2|s_0\rangle\langle s_0|.
\end{equation}
This operation flips the sign of $|s_0\rangle$ and keeps the component orthogonal to $|s_0\rangle$ unchanged. For large $N$, after $\sim \sqrt{N}$ calls of the oracle (followed by the diffusion operation after each oracle call), the state ends up in the target state $|\omega\rangle$ with nearly unit probability.

Grover's algorithm can be generalized to amplify the amplitudes of more than one target state, as described in the following. For an arbitrary state
\begin{equation}
    |s\rangle = \sum_{i=1}^N c_i |i\rangle,
\end{equation}
the task is to amplify the amplitude of all the basis states within a given subspace. Let $\hat{P}$ be the projector onto the target subspace, and $a$ be the ``weight'' of the initial state $|s\rangle$ in the target subspace, i.e., 
$$
a \equiv \langle s|\hat{P}|s\rangle.
$$

Similarly to the original Grover algorithm, the amplitude amplification implements the oracle and diffusion operators defined as
\begin{equation}
\begin{aligned}
   & \hat{O} \equiv I-2\hat{P},\\
   & \hat{D} \equiv I-2|s\rangle\langle s|.
\end{aligned}
\end{equation}
For large $N$, after $\sim 1/\sqrt{a}$ calls of the oracle and diffusion, the initial state $|s\rangle$ is projected to the target subspace with nearly unit probability. This approach, however, requires that the weight $a$ of the initial state in the target subspace is determined. In case $a$ is not known {\it a priori}, one can employ the amplitude estimation algorithm \cite{Brassard2000} first, and then apply the procedure described above.

If the task is not to find the projection of the initial state onto the target subspace, but to find a single basis state within the target subspace, as needed in the NNP1 protocol developed in the main text, the amplitude amplification algorithm can achieve this directly. That is, with (expected) $\sim 1/\sqrt{a}$ number of steps, one finds a single basis state within the desired subspace. The basic procedure of the algorithm is the following: with a fixed constant $1<c<2$, one should start with $l=0$ and compute $M = \lceil c^l \rceil$; Apply the oracle for a number of steps uniformly picked from $[1,M]$, and then measure the system. If a state within the target subspace is found, the algorithm terminates. Otherwise, increase $l$ by $1$ and repeat above. Proof of the algorithm can be found in Ref.~\cite{Brassard2000}.

\section{Robbins-Berry phase for spin 1}
\label{aBR}
To derive the Robbins-Berry phase
\cite{Robbins1994}, we consider a unit spin, $I=1$, in  an external field ${\bf b}(t)$ that changes with time adiabatically so that the initial and final field directions do not coincide but rather  differ by sign: ${\bf b}(t_{\rm in}) =-{\bf b}(t_{\rm fin}) = b\hat{z}$. Here, without loss of generality we assume that the initial field direction is along the $z$-axis.
Let the initial spin state $|0_z\rangle$ correspond to the zero spin projection on this axis.

Assume that during the adiabatic evolution, the magnetic field is always nonzero and the Hamiltonian is 
\begin{equation}
  H(t)={\bf b}(t) \cdot {\hat {\bf I}}.
  \label{hfield}
\end{equation}

Let  ${\bf b}=(b,  \theta, \varphi)$ be the parametrization of the field vector by the  time-dependent components in spherical coordinates, and 
\begin{equation}
R_{x}\left(\theta\right)=e^{-i\hat{I}_{x}\theta},\quad R_{z}\left(\varphi\right)=e^{-i\hat{I}_{z}\varphi}
\label{rotations}
\end{equation}
be the spin rotation operators.
The instantaneous eigenstates of the Hamiltonian (\ref{hfield}) are the spin projection states on the instantaneous field direction.
For the zero spin projection on the field axis this state is
\begin{equation}
  |0_{{\bm b}(t)}\rangle =R_z(\varphi) R_x(\theta) R_z^{-1}(\varphi) \left|0_z\right\rangle.
\label{eigenstate}
\end{equation}

The eigenvalues of $H$ are $-|b(t)|$, $0$, and $|b(t)|$, which are always separated by a finite gap from each other because ${\bf b}(t)$ is nonzero. 
According to the adiabatic theorem, the solution of the time-dependent Schr\"odinger equation in the adiabatic limit should coincide with $|0_{{\bm b}(t)}\rangle$ up to a phase factor $\exp\{i(\phi_d+\phi_{\rm geom})\}$, where 
$$
\phi_d =-\int_{T_{min}}^t d\tau\, \langle 0_{{\bm b}(\tau)}|H|0_{{\bm b}(\tau)}\rangle ,
$$
$$
\phi_{\rm geom}( C) =\int_{ C} {\bm A} ({\bf b}) \cdot\, d{\bf b}.
$$
Here, $C$ is the magnetic field trajectory,  and 
 $$
 {\bm A}({\bf b})\equiv i\langle 0_{\bf b}|\frac{\partial}{\partial{\bf b}}| 0_{\bf b } \rangle
 $$
 is  the standard Berry connection along this path.

The state $|0_{{\bm b}(t)}\rangle$ corresponds to the zero eigenvalue of $H$, so the dynamic phase is identically zero: $\phi_d=0$.
The explicit calculations of the Berry connection show that all its components, ${\bm A}=(A_b,A_{\theta},A_{\varphi})$ are identically zero, which means that the geometric phase correction to $|0_{{\bm b}(t)}\rangle$ is also identically zero.  Thus, $|0_{{\bm b}(t)}\rangle$ is the solution of the time-dependent Schr\"odinger equation with the Hamiltonian $H(t)$ in the adiabatic limit.

At the end of the evolution, ${\bf b}_{\rm fin}$ has opposite direction to the $z$ axis. Hence, the final state $|0_{fin}\rangle$ coincides with the initial state $|0_z\rangle$ up to an unknown phase factor that we now determine. The final state in  Eq.~(\ref{eigenstate}) corresponds  to $\theta =\pi$. Note also that $R_z^{-1}(\varphi) |0_z\rangle = |0_z\rangle $. Hence, the final state of the spin is given by
$$
|0_{\rm fin}\rangle = e^{i\pi \hat{I}_x}|0_z\rangle.
$$
The  phase difference between the initial and the final states is 
\begin{equation}
e^{i\phi} =\langle 0_z |0_{\rm fin}\rangle = \langle 0_z | \sum_{k=0}^{\infty} \frac{(i\pi \hat{I}_x)^k}{k!} |0_{z}\rangle,
\label{phase-prob1}
\end{equation}
which can be calculated by recalling the matrix form
$$
\hat{I}_x=\left(\begin{array}{ccc}
0&1/\sqrt{2} &0 \\
1/\sqrt{2} & 0&  1/\sqrt{2} \\
0&1/\sqrt{2} &0
\end{array}
\right).
$$
All odd powers of $I_x$ have zero expectation values over the state $|0_z\rangle$, whereas $\langle 0_z| \hat{I}_x^2 |0_z\rangle =1$, and $\hat{I}_x^4 =\hat{I}_x^2$. 
Then, the series in (\ref{phase-prob1}) can be summed as 
$$
e^{i\phi} =\sum_{k=0}^\infty \frac{(i\pi)^{2k}}{(2k)!} =\cos(\pi) =-1.  
$$
Thus, the accumulated phase by the end of the field sweep to the opposite direction is $\phi =\pi$. This is the Robbins-Berry   phase, which does not depend on the path of the field ${\bf b}(t)$ between its boundary values. 

 \section{Nonadiabatic transitions for spin-1 in time-dependent field}
\label{a-nonad}

The theory of nonadiabatic transitions for spin-1/2 in a time-dependent magnetic field is well established. Its generalization to problems with more than two interacting states remains an obscure topic but with some exceptions. Thus, in 1932, Majorana showed that any result for a spin-1/2 in a time-dependent magnetic field can be generalized to a spin of arbitrary size \cite{majorana}. Here, we review this generalization with application to our annealing problem for spin-1.

Consider again the Hamiltonian of a spin-1 in a time-dependent magnetic field:
\begin{equation}
H={\bf b}(t)\cdot \hat{{\bf I}},
\label{spin-1-mlz}
\end{equation}
and associate with it the Hamiltonians, $h_1$ and $h_2$, of two independent  spins-1/2 that are placed in the same as in  (\ref{spin-1-mlz}) time-dependent field, i.e.,  
\begin{equation}
h_{1} = h_2 = \frac{1}{2} {\bf b}(t)\cdot \hat{\bm \sigma}.
\label{H2-mlz}
\end{equation}
Note that $h_1$ and $h_2$ act in different spin spaces. Hence, both spins are described simultaneously by a combined Hamiltonian 
\begin{equation}
H'=h_1 \otimes {1}_2 + {1}_1  \otimes h_2,
\label{H3-mlz}
\end{equation}
where ${1}_{1,2}$ are  unit 2$\times$2 matrices acting in, respectively, the first and the second spin sectors. The Hamiltonian (\ref{H3-mlz}) is acting in space with four basis vectors: 
\begin{equation}
|1 \rangle \equiv | \uparrow \uparrow \rangle, \quad |-1 \rangle \equiv |\downarrow \downarrow \rangle, \quad |0\rangle \equiv \frac{1}{\sqrt{2}}\left( |\uparrow \downarrow \rangle +|\downarrow \uparrow \rangle \right),
\label{triplet-mlz}
\end{equation}
\begin{equation}
|-\rangle \equiv \frac{1}{\sqrt{2}}\left( |\uparrow \downarrow \rangle -|\downarrow \uparrow \rangle \right),
\label{singlet-mlz}
\end{equation}
where we use short notation $|\uparrow \uparrow \rangle \equiv |\uparrow \rangle \otimes |\uparrow \rangle$, e.t.c.. 
Since $|-\rangle$ is an eigenstate of $H'$ for all times, it decouples from the triplet (\ref{triplet-mlz}). Moreover, within the triplet (\ref{triplet-mlz}), $H'$ has the matrix form (\ref{spin-1-mlz}). Indeed, it is easy to check, e.g., that $\langle 1 |H' | 1\rangle=-\langle -1|H'|-1\rangle = b_z$, $\langle 1 | H' |0 \rangle =b_x/\sqrt{2}$, e.t.c.. 

Since spins-1/2 experience the same time-dependent field, their evolution over the time interval $t\in (T_{min},T_{max})$ is described by the same evolution matrix:
\begin{equation}
U_{1}=U_{2}=\left( \begin{array}{cc}
a & b\\
-b^* & a^*
\end{array} \right),
\label{u2-mlz}
\end{equation}
with complex amplitudes $a$ and $b$.
The evolution matrix for the Hamiltonian $H'$ factorizes as the direct product:
\begin{equation}
U'=U_1 \otimes U_2.
\label{uu-mlz}
\end{equation}
For example, if the initial state, at $t=T_{min}$, is $|1 \rangle =|\uparrow \uparrow \rangle$ then the amplitude of the state $|1\rangle $ at time $T_{max}$ is 
$$
\langle 1| U' |1 \rangle =a^2.
$$
Similarly, $\langle 0 |U'| 1\rangle=-\sqrt{2} ab^*$, whereas  $\langle 1| U'|-\rangle=0$, e.t.c.. Summarizing, if we know the evolution operator (\ref{u2-mlz}) for spin-1/2 in a time-dependent magnetic field, then we can also write the evolution matrix for the Hamiltonian that describes spin-1 in the same field:
\begin{equation}
U=\left( \begin{array}{ccc}
a^2 & \sqrt{2} ab & b^2 \\
-\sqrt{2} ab^* & |a|^2-|b|^2 & \sqrt{2} a^*b\\
(b^*)^2 &-\sqrt{2}a^*b^*  &(a^*)^2
\end{array}
\right).
\label{ufin-mlz}
\end{equation}

The central element, $U_{00}=|a^2|-|b^2|=2|a|^2-1$, of this matrix is the amplitude to stay on the zero-projection state after the evolution. Note that this element is purely real. For spin-1/2, the adiabatic evolution that flips the spin to the opposite direction corresponds to 
$|a|=0$ and $|b|=1$, which leads to $U_{00}=-1$, in agreement with  Robbins-Berry phase $\pi$ in Appendix~\ref{aBR}. Our result is more general: even in the case of small but finite nonadiabatic transitions, the element $U_{00}$ remains real and thus this $\pi$-phase is protected. 

For a quasi-adiabatic sweep of one magnetic field component from large negative to large positive values throughout an avoided crossing point, the probability of the nonadiabatic transition for spin-1/2 is generally given by the Dykhne formula \cite{fdykhne}:
\begin{equation}
    |a|^2=ce^{-2{\rm Im} \left[ \int_{0}^{t_0} d\tau \, \sqrt{{b_z^2+b_x(\tau)^2}} \right]},
    \label{Dykhne}
\end{equation}
where $t_0$ is the complex-valued time point that corresponds to closing the gap in the spectrum: $|b_z^2+b_x^2(t_0)|=0$. If there are many such points we should choose the one that minimizes the integral in (\ref{Dykhne}). Generally $c=1$, with exceptions in cases of rare symmetries. 

The Dykhne formula predicts an exponentially suppressed probability of a nonadiabatic transition $|a|^2\sim e^{-\eta \Delta T}$, where $\Delta$ is the minimal gap during the evolution and $T$ is a characteristic time of the transition through the avoided crossing; $\eta$ is a model-specific coefficient of order $1$. 
For our spin-1 models, the probability to make a nonadiabatic transition to the states with nonzero spin polarization on the final field axis is given by 
\begin{equation}
P_{ex}=1-|U_{00}|^2 \approx 4|a|^2.
    \label{nonad}
\end{equation}

%Finally, we estimate  this probability, using Eq.~(\ref{Dykhne}), for the protocols that encountered in the main text. For the Hamiltonian (\ref{ha2}) with the protocol (\ref{gt1}), the effective field components are $b_z=E_k$ and $b_x=\sinh(t/T)$. Thus the gap for avoided crossing is $\Delta =E_k$, and we expect that  $|a|^2\sim e^{-\eta E_kT}$ with $\eta=O(1)$.

%Since all nonzero $E_k$ are integers, the most dangerous situation can happen for the smallest possible gap $E_1=1$, for which
%$$
%\sqrt{b_z^2+b_x^2} = \cosh %(t/T).
%$$
%This field becomes zero for $t_k= i(\pi /2+\pi k)T$. The transition probability to the excited state is dominated by the point $t_0$ that is closest to the real time axis. 

%The integral in the exponent in (\ref{Dykhne}) is then
%$$
%-2 {\rm Im}  \int_{0}^{i\pi T/2}  \cosh (t/T) \, dt = -2T.
%$$
%The prefactor for this model is known to be $c=2$ \cite{fjoye}, so
%$$
%|a|^2 \approx 2e^{-2T}, 
%$$
%and for spin-1 sector with the minimal possible nonzero gap we find
%$$
%P_{\rm ex} \approx 8e^{-2T},
%$$ 
%which is exponentially small for $T\gg 1$, as we expected.

For the model (\ref{hag1}) with exponential coupling decay (\ref{expg1}), the invariant sectors have the field components $b_z=E_k-E$ and $b_x(t)=e^{-t/T}$. 
The Dykhne formula then predicts $|a|^2\approx e^{-\pi |E_k-E|T}$, and for spin-1 we find
$$
P_{ex}\approx 4e^{-\pi |E_k-E|T}.
$$